\documentclass[useAMS,usenatbib]{mn2e}
\usepackage{color,graphicx}
\usepackage{multirow}
\def\araa{ARA\&A}       	% Annual Review of Astron and Astrophys
\def\apj{ApJ}           	% Astrophysical Journal
\def\mnras{MNRAS}       	% Monthly Notices of the RAS
\def\gsim{\hspace{0.3em}\raisebox{0.4ex}{$>$}\hspace{-0.75em}\raisebox{-.7ex}{$\sim$}\hspace{0.3em}}

\title[SAI in protostellar discs]{Spiral arm instability - III. Fragmentation of primordial protostellar discs}
\author[S. Inoue\& N. Yoshida]
{\parbox[t]{\textwidth} 
{Shigeki Inoue$^{1,2,3,4}$\thanks{E-mail: shigeki.inoue@nao.ac.jp} \& Naoki Yoshida$^{3,4, 5}$}
\\ \\
$^{1}$Center for Computational Sciences, University of Tsukuba, Ten-nodai, 1-1-1 Tsukuba, Ibaraki 305-8577, Japan\\
$^{2}$Chile Observatory, National Astronomical Observatory of Japan, Mitaka, Tokyo 181-8588, Japan\\
$^{3}$Kavli Institute for the Physics and Mathematics of the Universe (WPI), UTIAS, The University of Tokyo, Chiba 277-8583, Japan\\
$^{4}$Department of Physics, School of Science, The University of Tokyo, Bunkyo, Tokyo 113-0033, Japan\\
$^{5}$Research Center for the Early Universe, School of Science, The University of Tokyo, Bunkyo, Tokyo 113-0033, Japan
}

\begin{document}

\pagerange{\pageref{firstpage}--\pageref{lastpage}} \pubyear{2014}

\maketitle

\label{firstpage}

% 250 words
\begin{abstract}
We study the gravitational instability and fragmentation of primordial protostellar discs by using high-resolution cosmological hydrodynamics simulations. We follow the formation and evolution of spiral arms in protostellar discs, examine the dynamical stability, and identify a physical mechanism of secondary protostar formation. We use linear perturbation theory based on the spiral-arm instability (SAI) analysis in our previous studies. We improve the analysis by incorporating the effects of finite thickness and shearing motion of arms, and derive the physical conditions for SAI in protostellar discs. Our analysis predicts accurately the stability and the onset of arm fragmentation that is determined by the balance between self-gravity and gas pressure plus the Coriolis force. Formation of secondary and multiple protostars in the discs is explained by the SAI, 
which is driven by self-gravity and thus can operate without rapid gas cooling. We can also predict the typical mass of the fragments, which is found to be in good agreement with the actual masses of secondary protostars formed in the simulation.
\end{abstract}


\section{Introduction}
\label{Intro}
The first generations of stars affects the thermal and chemical
state of the inter-galactic medium in the early Universe by photo-ionisation and heating, and 
by dispersing heavy elements synthesized within them. 
The first stars also affect the subsequent formation and evolution of galaxies 
and massive black holes through a variety of feedback effects \citep{BrommYoshida11}.
The initial stellar mass function is perhaps the most important quantity
that characterises the first stars and their overall impact to cosmic structure formation.
So far, primordial star formation in a cosmological context has been studied extensively, and 
detailed protostellar evolution calculations suggest that the first stars have a variety of masses with $10$--$1000~{\rm M_\odot}$ \citep[e.g.][]{hhy:14}. 

Formation of binaries or multiple stars has been also found in recent  
simulations that follow the
post-collapse evolution of primordial proto-stellar systems 
\citep{gbc:12, Stacy16, Susa19}.
A primordial proto-stellar disc is embedded within an accreting envelope \citep{Hosokawa16},
and often becomes unstable to fragment and yield multiple protostars, but
the exact mechanism(s) that drives the disc fragmentation has not been identified.
The frequency of disc fragmentation and the masses of the fragments are
critically important because low-mass primordial stars, if they are formed, survive until 
the present day and might be found in the local Universe. 

\citet{gbc:12} run cosmological simulations of primordial star formation to
show that multiple protostars are formed
in the circumstellar disc around the main protostar.
In their simulation, spiral arms are excited in the disc, which occasionally fragment to generate secondary protostars that have small masses. If the secondary stars are gravitationally bound within the disc, and do not get accreted by the central star, they will form a binary or multiple stellar system. It is important to understand the physical mechanism of disc fragmentation, and it appears that fragmentation of spiral arms is a critical step for the formation of multiple systems and star clusters. 

Fragmentation of a spiral arm is often discussed in terms of gravitational instability, such like non-linear evolution of Toomre-unstable regions \citep{s:60,t:64}. The Toomre analysis itself describes the linear stability of a two-dimensional region within a thin disc, and it remains unclear what is the end product of Toomre instability -- spiral arms or spherical clumps. 
Furthermore, the Toomre analysis is no longer applicable
if there is significant density contrast between perturbations (arms) 
and the disc.
Another important issue may be the effect of radiative cooling. It is often argued that rapid gas cooling is responsible for fragmentation of arms \citep[e.g.][]{g:01,rab:03}.
Since the gas in primordial protostellar discs evolve approximately isothermally, it is not clear
if fragmentation can be driven by radiative cooling. \citet{ttm:15} and \citet{tti:16} argue that rapid cooling is not essential for disc fragmentation.
While there are a variety of physical processes suggested, and
related criteria are proposed, the physical mechanism of primordial gas disc fragmentation has not been fully explored. 

In our previous studies \citep[][hereafter Papers I and II]{iy:18,iy:18b}, we present linear perturbation analysis for spiral arms in galactic discs and derive the physical conditions for spiral arm instability (SAI). We use the output of high-resolution simulations of isolated disc galaxies to show that SAI induce fragmentation of galactic spiral arms. We also derive the characteristic mass of the fragments formed via SAI. 
In this {\it Letter}, we apply the SAI analysis to protostellar discs formed in early primordial gas clouds. We investigate the physical mechanism of spiral-arm fragmentation through which secondary protostars are formed. 

\section{Numerical Simulations}
\label{sims}
We use the cosmological hydrodynamics simulations of \citet{gbc:12}, which are performed with the moving-mesh hydrodynamics code {\sc Arepo} \citep{arepo}. The simulations follow the collapse of primordial gas clouds in small-mass "minihaloes". Primordial gas chemistry and associated cooling processes are  followed for
the species of ${\rm H}$, ${\rm H^+}$, ${\rm H^-}$, ${\rm H_2^+}$, ${\rm H_2}$, ${\rm He}$, ${\rm He^+}$, ${\rm He^{++}}$, ${\rm D}$, ${\rm D^+}$, ${\rm HD}$ and free electrons. Details of the simulations are found in \citet{gbc:12}. We use their MH1 and MH4 runs. In the moving-mesh hydrodynamcis simulations, adoptive refinement is performed for gas cells such that a Jeans length is resolved with 32 cells.

\section{Instability Analysis}
\label{ana}
In Papers I and II, we derive the physical conditions for gravitational instability of a spiral arm
in a thin disc, based on linear analysis for azimuthal perturbations along the arm that 
is approximated to be tightly wound \citep[see also][]{tti:16}. The instability parameter for perturbations with wavenumber $k$ is given as 
\begin{equation}
S \equiv\frac{\sigma^2k^2 + \kappa^2}{\upi G~f(kW)\Upsilon k^2},
\label{Crit}
\end{equation}
where $G$ is the gravitational constant, $W$ and $\Upsilon$ are the half width and the line mass of the spiral arm defined as the mass per unit length, and $f(kW)\equiv[K_0(kW)L_{-1}(kW) + K_1(kW)L_0(kW)]$ with $K_i$ and $L_i$ denoting the modified Bessel and Struve functions of order $i$. We compute the epicyclic frequency $\kappa$ directly from the rotation velocity $v_{\phi}$ as
\begin{equation}
  \kappa^2=2\frac{v_{\phi}}{R}\left(\frac{\mathrm{d}v_{\phi}}{\mathrm{d}R} + \frac{v_{\phi}}{R}\right),
  \label{kappa}
\end{equation} 
as in \citet{idm:16}.
The velocity dispersion is calculated as $\sigma^2 = c_{\rm snd}^2+\sigma_\phi^2$, where $c_{\rm snd}$ and $\sigma_\phi$ are sound speed and the azimuthal turbulent velocity dispersion. We note that $\sigma$, $\Upsilon$, $W$ and $\kappa$ can be evaluated locally from the simulation output (see Section \ref{armwidth}). 

The instability parameter $S$ is expressed as a function of $k$. If $S (k)>1$ for all $k$, the spiral arm is stable against all perturbations. The instability condition for a spiral arm and for perturbations with $k$ is given by $\min [S (k)]<1$. Our previous analyses in Papers I,  II and \citet{tti:16} assume rigid rotation within the arm, i.e., $\kappa=2\Omega$, but here we improve the analysis by considering
the velocity shear.

\subsection{Thickness correction}
\label{thickness}
The above analysis using equation (\ref{Crit}) still ignores the vertical thickness of a spiral arm. Since we consider spiral arms excited in primordial protostellar discs, they may
have a large thickness to affect significantly the stability analysis. A thick arm has low mass-concentration, and the gravitational force on the disc plane is effectively reduced. Accordingly, using equation (\ref{Crit}) overestimates the strength of the gravitational instability. To take this effect into account, we adopt a correction proposed by \citet{t:64}, in which the arm has a vertical height of $h$ and density fluctuations within the arm are assumed to be independent of height from the disc plane. In this case, fluctuations of gravitational potential on the disc plane are weakened by a factor of $[1-\exp(-kh)]/kh$. With this thickness correction, equation (\ref{Crit}) is modified to
yield
\begin{equation}
S_{\rm T}\equiv\frac{(\sigma^2k^2 + \kappa^2)h}{\upi Gf(kW)\Upsilon k\left[1-\exp(-kh)\right]}.
\label{CritT}
\end{equation}
The instability condition for a spiral arm is given by $\min [S_{\rm T}(k)] < 1$. We define the vertical height of the arm as $h\equiv(\sigma_z^2+c_{\rm snd}^2)/(\upi G\Sigma)$, where $\sigma_z$ is the vertical velocity dispersion \citep{e:11}. Although there are other studies that propose slightly different correction factors \citep[e.g.][]{gl:65,v:70,r:92}, they give essentially the same instability parameter\footnote{These analyses consider the instability of local disc regions against radial perturbations, rather than that of a spiral arm against azimuthal ones. However, the finite thickness only affects the estimate of the strength of gravity on the disc plane, and the instability analysis does not differ between radial and azimuthal perturbations as long as the direction of the perturbations is independent of time \citep{t:64}.}. 

\subsection{The arm width and epicyclic frequency}
\label{armwidth}
To compute the instability parameter $S_{\rm T}$, we use vertically averaged values of the physical quantities such as $\Sigma$, $\sigma_\phi$, $\sigma_z$, $c_{\rm snd}$ and $v_{\phi}$. In practice, we first apply two-dimensional Gaussian smoothing with $r_{\rm smooth} = 0.3~{\rm AU}$ to the gas distribution. Then we make polar plots of the quantities as functions of $(R,\phi)$ for each snapshot.

\begin{figure*}
  \includegraphics[bb=0 0 1764 923, width=\hsize]{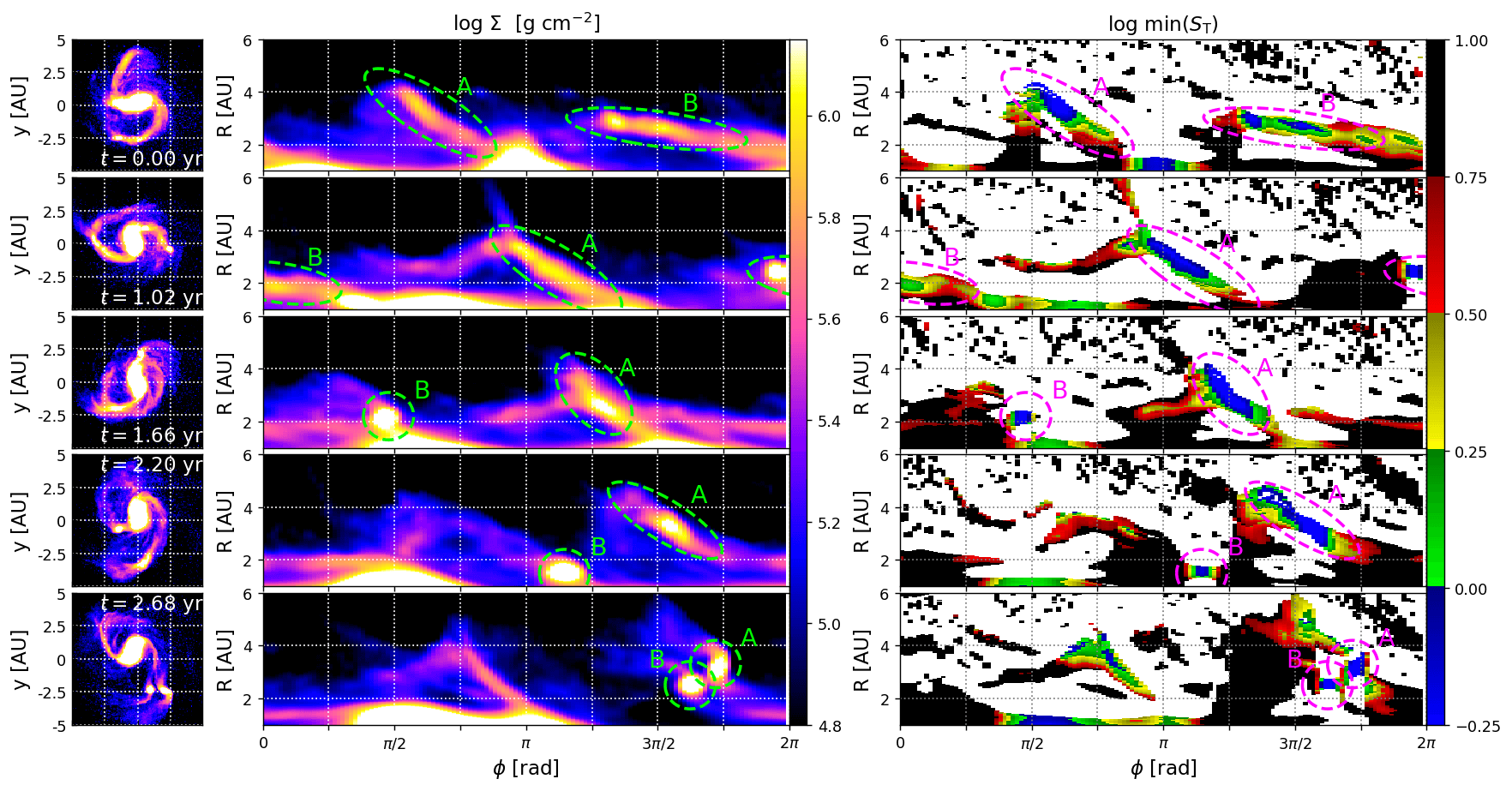}
  \caption{The polar-map representation of our analysis for the run that shows arm fragmentation. The left and central panels show the face-on gas distributions in the Cartesian and polar coordinates. Times elapsed since the first snapshot (top panel) are indicated in the left panels. The right panels show the thickness-corrected instability parameters, $\min(S_{\rm T})$, where unstable regions with $\min(S_{\rm T})<1$ are coloured in blue. In the central and right panels, two fragmenting arms are marked with dashed ellipses and labelled as `A' and `B'.}
  \label{mh4}
\end{figure*}

In order to examine the SAI, we first need to detect spiral arms and measure their half widths $W$. 
To this end, we perform one-dimensional Gaussian fitting along the radial direction at a given $\phi$ using the polar map of the surface density $\Sigma(R,\phi)$.
The fitting function is defined as $\tilde{\Sigma}(R,\xi,\phi)=\Sigma(R,\phi)\exp[-\xi^2/2w^2]$, where $\xi$ represents radial offset from $R$.
%, and $\Sigma(R,\phi)$ is an actual surface density in the %simulation. 
The fitting is performed in the range of $-1.55w<\xi<1.55w$ with varying $w$. We search for $w$ that minimises the goodness-of-fit $\chi^2$. Then the arm half width is given by $W=1.55w$. With this definition of $W$, the estimated Gaussian surface density at the edge of the spiral arm corresponds to 30 percent of the peak value \citep{tti:16}, i.e., $\tilde{\Sigma}(R,\pm W,\phi)=0.3\Sigma(R,\phi)$. The line-mass of an arm is obtained as $\Upsilon(R,\phi)=1.44W\Sigma$ (see Paper I). 
%If there is a crest of the arm at $R$ and the density distribution is nearly %Gaussian, $\chi^2$ becomes significantly lower than unity. 
We define spiral arms to be regions where $\log\chi^2<-0.25$. Although this threshold is arbitrary, our subsequent procedure to compute $S_{\rm T}$ is independent of this choice. Note that the above procedures assume a spiral arm to have a Gaussian density distribution and a pitch angle of $\theta=0$. Our Papers I and II confirm that this assumption nevertheless allows us to detect spiral arms robustly and to derive the
instability parameter accurately.

We compute the epicyclic frequency $\kappa$ in the following manner. The rotation velocity $v_{\phi}(R,\phi)$ is calculated from the smoothed velocity fields. The local velocity gradient $\mathrm{d}v_{\phi}/\mathrm{d}R$ at $(R,\phi)$ is computed by linear least-square fitting for $v_{\phi}$ in the radial range of $-W<\xi<W$, and $\kappa(R,\phi)$ is computed from equation (\ref{kappa}). All the quantities necessary for computing $S$ are obtained as functions of $(R,\phi)$. 

\section{Results}
\label{result}
Spiral arms are excited in protostellar discs in all our simulations. Although some arms occasionally fragment to form secondary protostars, not all the spiral arms do so. As illustrative cases, we focus on two characteristic
runs; one shows arm fragmentation and the other with stable arms. We apply our analysis described in Section \ref{ana} and show that the fragmentation in the former case is explained by SAI.

\subsection{Fragmentation and multiple protostars}
Fig. \ref{mh4} shows the results of our polar-map analysis for the run where two spiral arms fragment to yield secondary protostars. The central star has a mass of $2.13~{\rm M_\odot}$ within $1.5~{\rm AU}$.\footnote{At $t=2.68~{\rm yr}$, the spiral arms are located from $R\simeq1.5~{\rm AU}$ from the central star. We define this radius to be the outer boundary of the main star in this run. Note that this definition is different from that of \citet{gbc:12}.} We find that the spiral arms and the inter-arm regions have similar temperatures $T\simeq1500$--$2000~{\rm K}$, except in shock-heated regions with $T\gsim3000~{\rm K}$. Using the cooling rates of a primordial gas computed by \citet{o:01}, we calculate the characteristic cooling time in the arms to be $\tau_{\rm cool} \sim 1 ~{\rm year}$. The ratio of the cooling time to the disc rotation is then $\tau_{\rm cool}\,\Omega\sim1$ within the arms; the cooling time is comparable to, but not significantly smaller than the disc rotation timescale. The unstable, fragmenting parts are marked with dashed ellipses labelled as Arms A and B. 
%The left and central panels show the face-on surface densities of gas in the Cartesian and polar %coordinates, respectively. 
We define the time of the first snapshot (top panels) to be $t=0$, when a pair of prominent spiral arms have developed. In the next few years, both the arms fragment to yield two secondary stars. Arm A has initially a smooth density distribution, and the density increases until $t=1.66~{\rm yr}$ (the third snapshot). Finally at $t=2.20~{\rm yr}$, Arm A fragments to form a secondary star. Similarly, Arm B bears another secondary star at $t=1.66~{\rm yr}$.

The right panels of Fig. \ref{mh4} indicate the instability parameters $\min(S_{\rm T})$ in the spiral-arm regions with $\log\chi^2<-0.25$ (see Section \ref{armwidth}), where the blue regions with $\min (S_{\rm T}) < 1$ are regarded as unstable. The regions indeed fragment by $t=2.68$ yr.
Our SAI analysis accurately predicts the fragmentation in the simulation, and can also quantify the degree of gravitational instability by the value of $\min (S_{\rm T})$.\footnote{The region at $(R,\phi)\simeq(1~{\rm AU}, 3~{\rm rad})$ also indicate $\min (S_{\rm T})<1$ at $t=0$. However, this region corresponds to the tip of the elongated envelope of the main star. Our analysis assuming an anulus is not applicable to such a bar structure.} This suggests that the physical mechanism of the spiral-arm fragmentation is linear dynamical instability driven by self-gravity of the arm.

In Paper I, we successfully predict the characteristic mass of clumps formed by SAI. Here, we aim at predicting the mass of fragments (protostars) in the disc by following essentially the same procedure. The wavelength of the most unstable perturbation can be obtained as $\lambda_{\rm MU}=2\upi/k_{\rm MU}$ where $k_{\rm MU}$ is the wavenumber that gives the minimum $S_{\rm T}(k)$. In the snapshot at $t=0$ in Fig. \ref{mh4}, for both the two arms, we find that $\lambda_{\rm MU}\simeq 3~{\rm AU}$ and the arm widths to be $W\simeq 0.5~{\rm AU}$. In this case ($\lambda_{\rm MU}\gsim2W$), the most unstable perturbation with $\lambda_{\rm MU}$ is expected to collapse in the direction along the spiral arm. Then, the mass of a collapsing clump is estimated to be $M_{\rm cl}\sim\Upsilon~\lambda_{\rm MU}$. The masses of the fragments (secondary stars) can be already predicted at $t=0$ to be $M_{\rm cl}\simeq0.3$ and $0.2~{\rm M_\odot}$ for Arms A and B, respectively. In the snapshot at $t=2.68~{\rm yr}$ (Fig. \ref{mh4}), we find that the actual masses of the simulated secondary stars are $0.16$ and $0.15~{\rm M_\odot}$ for A and B\footnote{In the mass estimations of the secondary stars, we define their clump centres to be the positions of the highest $\Sigma$. The masses of the secondary stars are computed as the enclosed masses within $0.5~{\rm AU}$ from their centres in the projected density maps, where the separation between the two secondary stars is $\simeq1~{\rm AU}$.}, which are in good agreement with the estimated values. 

We note that the regions with $\min (S_{\rm T})<1$ in Arm B are separated into the two segments at $t=0$, both of which are already shorter than $\lambda_{\rm MU}\simeq 3~{\rm AU}$. If we consider that a larger region with $\min (S_{\rm T})\sim 1$ between the two segments is also marginally unstable, the collapsing part in Arm B extends approximately over $3~{\rm AU}$. The entire region soon collapses to a single secondary star. The presence of the region with $\min (S_{\rm T})\sim 1$ within the collapsing arm may imply the limit of accuracy of our SAI analysis and/or properly analysing non-linear regime of the spiral arm fragmentation.

Overall, our SAI analysis for this simulation suggests that the arm fragmentation is driven by self-gravity that exceeds the stabilising effect by gas pressure and Coriolis force. The secondary protostars in A and B keep roughly constant masses until $t=2.68~{\rm yr}$, but afterwards they strongly interact with each other and exchange their masses.

\subsection{Stable arms}
\begin{figure}
  \includegraphics[bb=0 0 903 1183, width=\hsize]{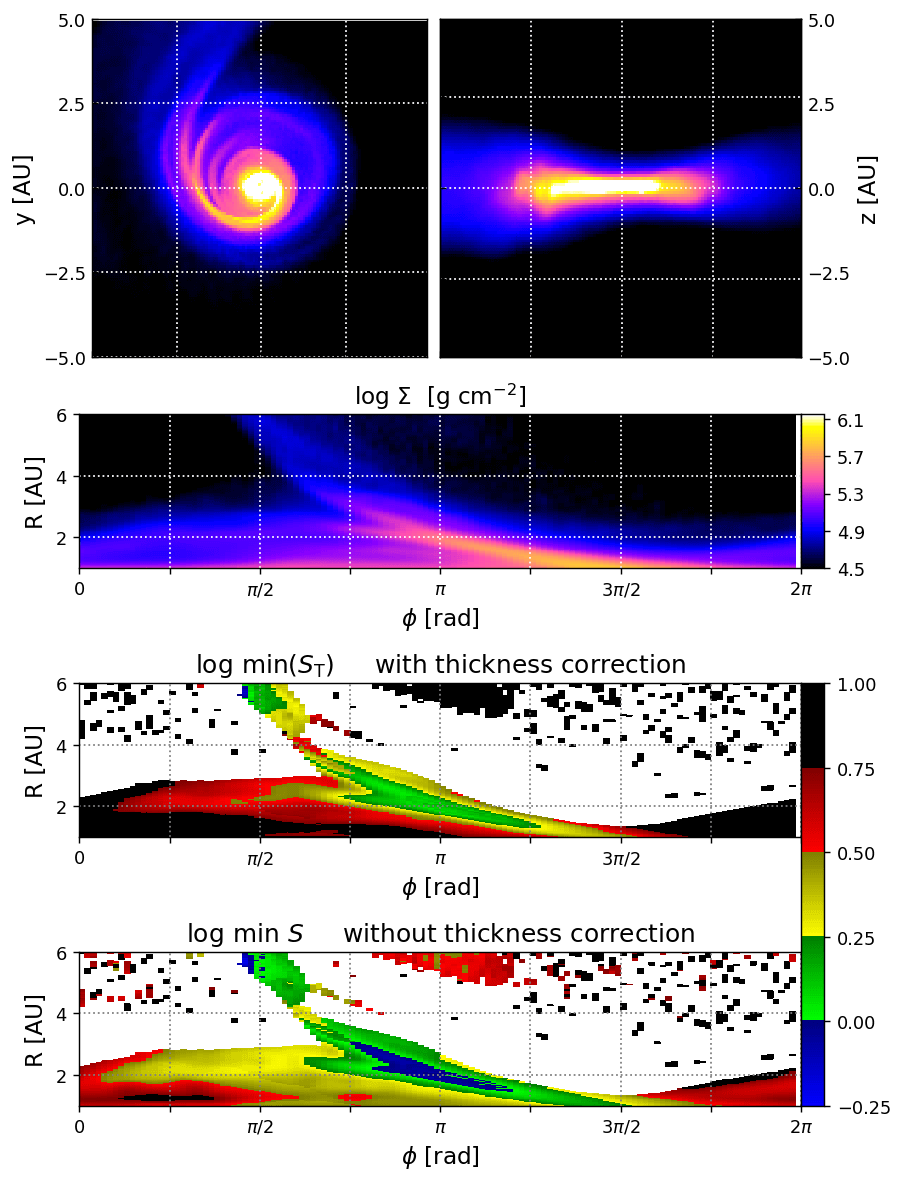}
  \caption{Our polar-map analysis for the non-fragmenting case. The top panels show gas distributions in the face-on (left) and edge-on (right) views. The second panel from the top shows the face-on density map but in the polar coordinates. The bottom two panels indicate the instability parameters with and without the thickness correction, i.e., $\min S_{\rm T}$ and $\min S$.}
  \label{mh1}
\end{figure}
Fig. \ref{mh1} shows the result for another run in which spiral arms do not fragment. The central star has a mass of $1.10~{\rm M_\odot}$ within $0.5~{\rm AU}$. The disc has nearly uniform temperatures $T\simeq1500$--$2000~{\rm K}$. Although we find $\tau_{\rm cool}\,\Omega\sim1$ within the arm, it does not fragment. This is interesting because we have shown that the fragmenting arm analysed in the previous section has also $\tau_{\rm cool}\,\Omega\sim1$. It appears that the cooling criterion of Gammie (2001) is not readily applicable to the fragmentation
of the spiral arms. 
Our SAI analysis gives $\min (S_{\rm T})>1$ at all the plotted region, suggesting that the arms {\it should} be stable. 
The disc has a large thickness
as can be seen in the edge-on view, 
where we also see significant flaring around the position of the spiral arm. 
Comparing the bottom two panels, we find that the thickness correction is critically important. Without the correction, there are portions with $\min (S) <1$ within the arm, suggesting that 
the arm should fragment, contrary to what is observed in the following evolution. Thick discs are often found in the accreting protostars such as 
those in our cosmological simulations. We conclude that the simple analysis with a razor-thin disc approximation overestimates the instability and thus erroneously predicts instability and fragmentation.
We also note that the often used
Toomre criterion $Q < 1$ is satisfied on the arm at the time shown in 
Fig. \ref{mh1}, but no fragmentation is seen in the subsequent evolution. 
This again supports our argument that Toomre instability
is not the direct cause of secondary protostar formation. 
    
\section{Discussion and conclusions}
\label{discussion}
We have successfully applied the SAI analysis developed in our Paper I/II to primordial protostellar discs.
Our improved analysis with finite
thickness correction and shearing motions accurately
describe the evolution and stability of spiral arms.
The physical process of secondary protostar formation can be summarized as follows.
First, fragmentation is triggered by gravitational instability of the arms, rather than by non-linear processes following the two-dimensional collapse owing to Toomre instability. Second, fragmentation is not driven by gas cooling but by self-gravity of the arm. When the self-gravity exceeds the stabilizing effects due to gas pressure and the Coriolis force, the spiral arm is subject to fragment and bears a secondary protostar. 

We confirm a two-step process for secondary protostar formation as suggested by \cite{tti:16}. Namely, the circumstellar disc around a primordial protostar can excite spiral arms through Toomre instability and/or swing amplification, and then SAI operates in some of the spiral arms when they satisfy $\min(S_{\rm T})<1$. 

The consistency between our linear analysis and the direct simulation result suggests that the fragmentation of arms can occur even if gas cooling is not efficient. \citet{g:01} argues that Toomre instability does not cause fragmentation since the instability also induce turbulent motions that recover a stable state by effectively heating up the disc. Instead, rapid 
gas cooling is thought to suppress the turbulent generation and thus to trigger fragmentation. In such a case, the instability condition is simply given by the ratio of the local cooling time scale to the disc rotation time. We find that both discs in our simulations marginally satisfy the Gammie's criterion. However, fragmentation occurs only in one disc, as our SAI analysis correctly predict. This implies that the rapid cooling is not a sufficient condition. Our analysis adopts a barotropic equation of state for gas, and thus the instability condition of $\min (S_{\rm T})<1$ is not directly related to the radiative cooling rate \citep[see also][]{tti:16}. Furthermore, SAI can operate
in stellar arms in $N$-body simulations where there is no dissipation (Paper I). With this result, we therefore conclude that rapid cooling is neither sufficient nor necessary conditions for fragmentation in protostellar discs. 

From the instability parameter given by equation (\ref{CritT}), it is expected that SAI tends to operate in massive (large $\Upsilon$), thin (small $W$ and $h$) and cold (small $c_{\rm snd}$ and $\sigma_\phi$) spiral arms around a less massive central star (small $\kappa$). Therefore, we expect SAI-driven fragmentation occurs preferentially in the early
stages of protostellar evolution when the gas mass accretion rate is large \citep{Yoshida08, Hosokawa16}.
Further studies are clearly warranted to investigate the formation of protostars via arm/disc fragmentation during later evolutionary stages of protostellar systems.

Formation of binary or multiple stars has been found in a number of numerical simulations \citep[e.g.][]{Clark11, Stacy14} 
but the exact reason for the disc fragmentation has not been identified. 
Often, complexities in technical implementation of 
sink particle generation and star-gas interaction
hampered proper understanding of the physical mechanism.
We have shown, for the first time, that 
the SAI drives the formation of secondary protostars. 
It is important to explore other mechanisms such as break-up or fission of protostars \citep{Tohline02} 
and filament fragmentation \citep{Chiaki16}.

\section*{Acknowledgements}
We are grateful to the reviewer for his/her useful comments. We thank Volker Springel for kindly providing the simulation code {\sc Arepo} and Thomas Greif for providing the outputs of his simulations. This study was supported by World Premier International Research Center Initiative (WPI), MEXT, Japan and by SPPEXA through JST CREST JPMHCR1414. SI receives the funding from KAKENHI Grant-in-Aid for Young Scientists (B), No. 17K17677. 
\vspace{0.5\baselineskip}

\noindent
Final note: This paper is dedicated to our friend and colleague, Dr. Thomas H. Greif (1981-2019).

%\bibliography{references}

\end{document}